\documentstyle[prl,aps,multicol,epsfig]{revtex}
\begin{document} 
\draft 
\title{Fermiology of a 1D heavy electron metal}

\author{C. Gr\"ober and R. Eder}
\address{Institut f\"ur Theoretische Physik, Universit\"at W\"urzburg,
Am Hubland,  97074 W\"urzburg, Germany}
\date{\today}
\maketitle

\begin{abstract}
We present a Quantum Monte Carlo (QMC) study of the 1D Kondo lattice at 
non-integer filling, i.e. a one-dimensional version of a Heavy Fermion metal. 
For this special system the minus-sign problem turns out to be greatly
reduced, so that accurate QMC simulations for temperatures as low as 1\% of the
conduction electron bandwidth are feasible. The single particle Green's 
function shows and intricate network of low-energy bands at low temperature, 
with a Fermi surface volume that comprises both $c$ and $f$-electrons. As the 
temperature increases the system evolves through two distinct crossover 
temperatures into a very simple band structure with one free $c$-electron band
and a disconnected upper and lower Hubbard band for $f$-electrons. The $f$ 
electrons thus drop out of the Fermi surface volume as the temperature 
increases.
\end{abstract}
\begin{figure}
\epsfxsize=6.0cm
\vspace{-0.0cm}
\hspace{-0.0cm}\epsfig{file=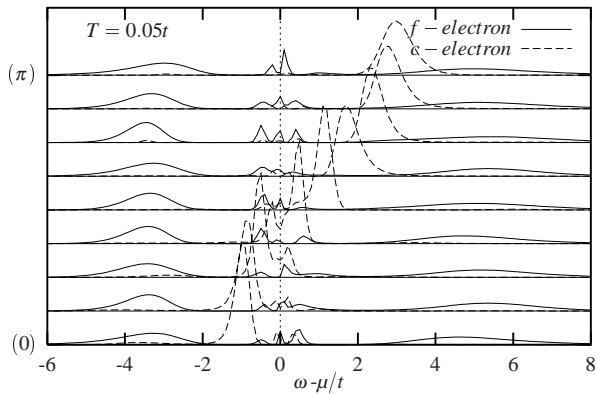,width=8.0cm,angle=0.0}
\vspace{0.0cm}
\narrowtext
\caption[]{Single particle spectral function for the
$16$-unit cell system.}
\label{fig1} 
\end{figure}

\begin{figure}
\epsfxsize=6.0cm
\vspace{-0.0cm}
\hspace{0.5cm}\epsfig{file=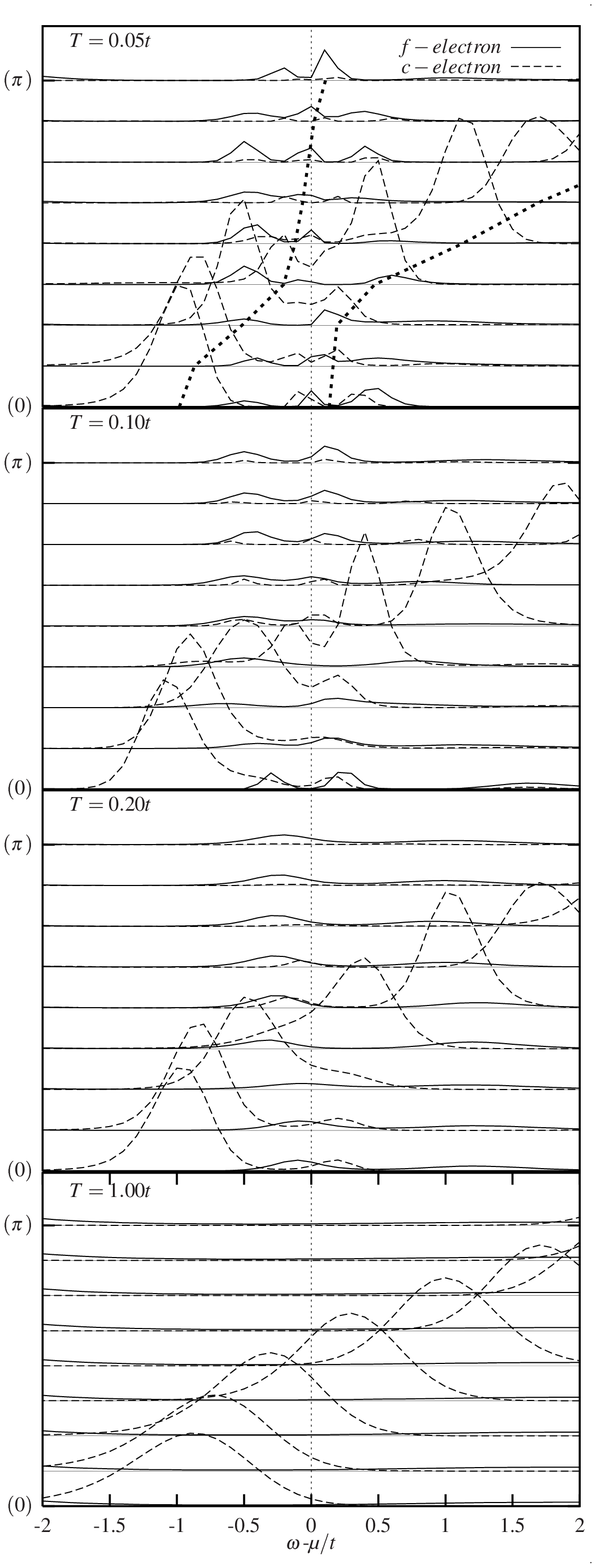,width=7.0cm,angle=0.0}
\vspace{0.0cm}
\narrowtext
\caption[]{Single particle spectral function
at different temperatures. The thick dotted line
for $T$$=$$0.05t$
shows a plausible form of the low-temperature band structure.}
\label{fig2} 
\end{figure}

\begin{figure}
\epsfxsize=6.0cm
\vspace{-0.0cm}
\hspace{0.5cm}\epsfig{file=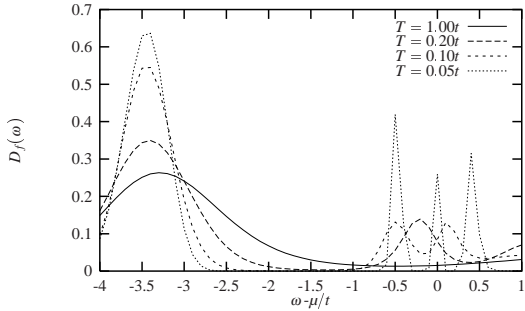,width=7.0cm,angle=0.0}
\vspace{0.0cm}
\narrowtext
\caption[]{Angle-integrated $f$-electron spectral weight.}
\label{fig3} 
\end{figure}

 
\begin{multicols}{2}
The electronic structure of heavy electron compounds remains
a largely unresolved problem of solid state 
physics\cite{Stewart,Fulde}. These materials show two
characteristic `crossovers temperatures' in their transport properties, 
which are 
commonly called the Kondo temperature, $T_{K}$,
and the coherence temperature, $T_{coh}$$<$$T_{K}$.
At temperatures $<T_{coh}$ the materials are Fermi liquids,
whereby the strongly correlated and almost
localized $f$-electrons do participate
in the Fermi surface volume. This has been established
by a wide variety of de-Haas van-Alphen (dHvA) 
experiments\cite{lonzarich,dhva}.
That the Fermi surface volume in heavy electron metals may be subject
to drastic changes, however, has been demonstrated by dHvA experiments 
in the neighborhood of the so-called
metamagnetic transition\cite{Aoki}. This transition appears to correspond to 
the $f$-electrons `dropping out' of the Fermi surface volume.
While the requirement of long quasiparticle lifetime
restricts dHvA experiments to very low temperatures and thus does not usually
allow to scan the temperature evolution of the Fermi surface,
it has been conjectured\cite{Fuldebuch} that the $f$-electrons do also
`drop out' of the Fermi surface volume as the temperature
increases above $T_{coh}$. The observation that the
iso-structural compounds CeRu$_2$Si$_2$ and CeRu$_2$Ge$_2$
have Fermi surfaces, which differ rather precisely by the volume
associated with the single 
$f$-electron/unit cell\cite{lonzarich,Fuldebuch,kinglonzarich}
would be quite consistent with
this scenario if one were to assume\cite{Fuldebuch,kinglonzarich}
that $T_{coh}$ is higher than the
experimental temperature for CeRu$_2$Si$_2$ but lower for CeRu$_2$Ge$_2$.
More precise information about the temperature-dependence
of the electronic structure of the Kondo lattice therefore would be
desirable. In the following we want to present information
from computer simulations. We have performed a Quantum Monte Carlo (QMC)
study for a simple 1D tight-binding version of the Kondo lattice
(or periodic Anderson model), with Hamiltonian
\begin{eqnarray}
H &=& 
-t\sum_{i,\sigma} 
(c_{i+1,\sigma}^\dagger 
c_{i,\sigma}^{}  + H.c.)
- V \sum_{i,\sigma} (c_{i,\sigma}^\dagger f_{i,\sigma}^{} + H.c.)
\nonumber \\
&-& \epsilon_f \sum_{i,\sigma} n_{i,\sigma}
+ U\sum_{i} f_{i,\uparrow}^\dagger f_{i,\uparrow}^{}
 f_{i,\downarrow}^\dagger f_{i,\downarrow}^{}.
\label{kondo1}
\end{eqnarray}
Here $c_{i,\sigma}^\dagger$  ($f_{i,\sigma}^\dagger$)
creates a conduction electron ($f$-electron)
in cell $i$, $n_{i,\sigma}$$=$$f_{i,\sigma}^\dagger f_{i,\sigma}^{}$.
In the case of `half-filling'
i.e. two electrons/unit cell and in the so-called
symmetric case, $\epsilon_f$$=$$U/2$, the model has particle-hole symmetry
whence the minus-sign problem of the QMC procedure
is absent and very low temperatures can be reached\cite{koiso}. 
Surprisingly enough we found that this highly desirable feature
persists to a high degree also away from half filling. For example
even at temperature $T$$=$$0.05t$, corresponding to
$1.2$\% of the conduction electron bandwidth, we obtained a QMC expectation 
value for the Fermionic sign of $0.605$ and thus
obtain very well-converged spectra.
This allowed to perform a detailed temperature scan of the
entire dynamics of the model. In the following, we use the
parameters $U$$=$$2\epsilon_f$$=$$8t$, $V$$=$$t$. Throughout
we have fixed the chemical potential at $\mu$$=$$-t$; this gives an electron 
density of $n$$=$$1.6$, with variations of less than $1$\% as the
temperature changes from $0.03t$ to $t$. The `all electron Fermi momentum'
$k_F$ then is $0.8\pi$, the `frozen core Fermi momentum' obtained
for unhybridized conduction electrons (i.e. $V$$=$$0$) would 
be $k_F^0$$=$$0.3\pi$.\\
Figure \ref{fig1} shows the 
single particle spectral function at the lowest temperature 
for which it was calculated,
$T$$=$$0.05t$. Thereby two different spectra are shown, one of them
for electron creation/annihilation in the $c$-orbitals, the
other one for $f$-orbitals.
The spectra contain several distinct features, which
are consistent with previous
exact diagonalization\cite{tsutsui} and analytical\cite{eos} results:
at the highest excitation energies, we have two dispersionless
$f$-like Hubbard bands, separated by an energy of $\approx U$.
In between these, a $c$-like
tight-binding band with dispersion $\epsilon_{\bbox{k}}$$=$$ -2t\cos(k)$
can be seen. This band seems to cross the chemical potential
close to $k_F^0$.
Closer inspection of the spectra shows, however, that this
is not the true Fermi momentum of the system. Very close
to the chemical potential (see Figure
\ref{fig2} for a close-up of 
\noindent 
this region) there is a rather intricate
network of $f$-like bands with low spectral weight.
Some of these bands apparently hybridize with
the $c$-like band. It should be noted that the peak positions of the
$c$- and $f$-like band do not always agree
perfectly, but actually the
MaxEnt procedure reconstructs the spectral densities
of the two species separately, and does not use the
peak positions of the other species as input information.
The possible discrepancies between $c$-like and $f$-like peak positions
thus give a feeling for the accuracy of the results.
For most momenta a total of three $f$-like bands can be distinguished.
In the outer half of the Brillouin zone
there is one band right at $\mu$, which
seems to disperse through $\mu$ approximately at
$k_F$, the Fermi momentum for $c$ {\em and} $f$-electrons combined.
This band seems to hybridize with the $c$-like cosine-band,
producing the familiar hybridization band structure indicated
in the Figure. In addition there are  two $f$-like sidebands
at slightly higher excitation energies $\approx \pm0.5t$.
While we cannot rule out
that these side-bands do have a tiny dispersion, they never approach the 
chemical potential. The three-band structure around $\mu$
can also be seen very clearly in the
angle-integrated $f$-electron spectral density, see Figure
\ref{fig3}. The angle-integrated
spectrum is actually {\em not} obtained by averaging over the
angle resolved spectra. Rather this quantity is obtained by an
independent measurement and MaxEnt run, and thus might be viewed
as an independent cross check for the band structure.
All in all the low energy band structure is 
quite reminiscent of our earlier results for the Kondo insulator\cite{koiso},
particularly the $f$-like
side-bands could also be clearly identified in this case.
With the exception of the $f$-like sidebands
one can describe the low-energy band structure phenomenologically
by the hybridization of the $c$-like cosine band with
a practically dispersionless `effective' $f$-level, which is
pinned to the chemical potential of the {\em unhybridized}
conduction electrons. This is precisely the picture inferred
from previous exact diagonalization\cite{tsutsui} 
and analytical\cite{eos} calculations.\\
We proceed to a discussion of the temperature dependence of the
band structure around $\mu$. As the temperature is increased to
$T$$=$$0.1t$, the $f$-like bands near $\mu$ generally become
quite broad. Near $k$$=$$\pi$ the upper 
\noindent 
side-band becomes very diffuse and shifted toward higher
energies $1.5t$.
The lower side band still can be clearly distinguished,
and also near $k$$=$$0$ an $f$-like part slightly above
$\mu$ is clearly resolved. This can also be seen in the
angle-integrated $f$-density (Figure \ref{fig3}): the lower side-band and the
band at $\mu$ persist, but the upper side-band is shifted to
relative large energies. In between these two bands there is still
an indication of the dispersive $f$-band which crosses $\mu$.
This changes completely at $T$$=$$0.2t$,
\noindent 
where the band crossing through
$\mu$ has disappeared and only the side-bands remains. 
The lower one
seems to have shifted to somewhat lower excitation energy, but does not
reach $\mu$, see also Figure \ref{fig3}.
The only dispersive band which could
intersect $\mu$ now is the $c$-like conduction band, which
crosses $\mu$ near $k_F^0$. The low-energy $f$-like band
which intersected $\mu$ at $k_F$
has disappeared, and the only $f$-like feature near $\mu$ are
the two side-bands. These are practically dispersionless and never
even approach $\mu$, so that there is no more
band which could cross $\mu$ at the `all-electron Luttinger momentum'
$k_F$. This suggests that at this temperature the
Fermi surface is purely $c$-like, with the smaller Fermi momentum $k_F^0$:
the $f$-electrons have dropped out of the Fermi surface volume.
Precisely the same behavior has previously been 
observed in the Kondo insulator: upon increasing $T$
the $f$-like bands which formed the edge of the
insulating gap at low temperatures disappeared, the
dispersionless side-bands persist, and the
Fermi surface becomes $c$-like. The only difference is that in the case of the
Kondo insulator the collapse of the
Fermi surface volume is equivalent to a 
insulator-to-metal transition\cite{koiso}.
Proceeding to the highest temperature we have studied, $T$$=$$t$, we note
that there even the two $f$-like sidebands have disappeared and the
Fermi surface now is obviously purely $c$-like, with
Fermi momentum $k_F^0$. The angle-integrated $f$-electron density
shows no more spectral weight at $\mu$, the only remaining $f$-like
feature on the photoemission side is the lower Hubbard band at
energy $\approx -3.2t$. Increasing the temperature thus leads to a
transfer of $f$-like weight from the Fermi energy to the
Hubbard band.  With the exception of the side bands
the overall shape of the $f$-density
at low temperatures is also quite consistent with
the results from $1/N$ expansion for the
Anderson impurity model\cite{Gunnarson}. Also, the overall
evolution with temperature, including the disappearance of
all $f$-like weight from the region around $\mu$ at high
temperature is completely consistent with
the situation for the Kondo insulator\cite{koiso}.\\
To get a more detailed picture of the $T$-dependence of the
Fermiology, we consider the momentum distribution
$n_\alpha(\bbox{k})$$=$$\sum_\sigma
\langle \alpha_{\bbox{k}\sigma}^\dagger
\alpha_{\bbox{k}\sigma} \rangle$, where $\alpha$$=$$c,f$.
We focus on the $f$-like distribution and start out
at the lowest temperature. 
\begin{figure}
\epsfxsize=6.0cm
\vspace{-0.5cm}
\hspace{-0.5cm}\epsfig{file=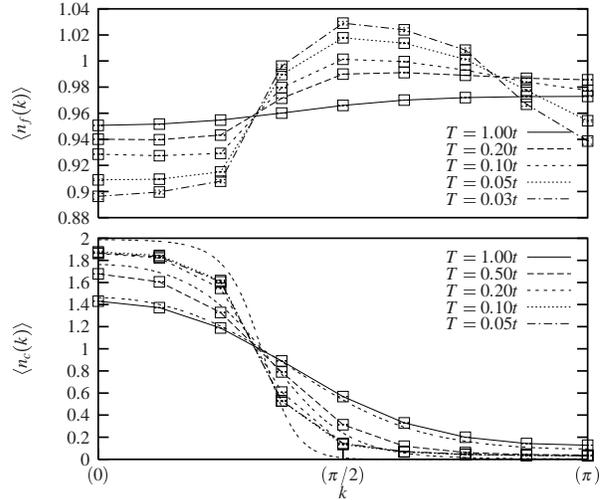,width=8.0cm,angle=0.0}
\vspace{0.0cm}
\narrowtext
\caption[]{Momentum distributions at
different temperatures. The thin dashed lines give the
Fermi-Dirac distribution computed with 
$\epsilon_k=-2t\cos(k)$ and $\mu=-t$
at $T=t, 0.5t, 0.2t$.}
\label{fig4} 
\end{figure}
\noindent 
As $k$ increases 
from the center of the Brillouin zone, there is
first of all a steep rise of $n_f(k)$ at approximately $k_F^0$. Inspection of
the single particle spectrum (Figure \ref{fig2})
reveals that this is due to an increase
of the spectral weight in the side-band for $\omega-\mu<0$ -
note that $n_f(k) $$=$$ \int d\omega A_f(k,\omega) f(\omega)$,
where $f(\omega)$ is the Fermi function.
As we approach $k$$=$$\pi$ and pass through $k_F$
the $f$-distribution drops again,
which this time is associated with the crossing of the central 
$f$-like band through $\mu$ (see Figure \ref{fig2}). 
Due to the flatness of this central
band and its small spectral weight per $k$-point, $Z$, the
`drop' is actually very shallow; in a Fermi liquid
the momentum
distribution near $k_F$ can be written as
\[
n(k) = const + \frac{Zv_F}{k_B T}(k-k_F),
\]
where $Z$ is the quasiparticle weight, $v_F$ the Fermi velocity
and $const$ the contribution from the incoherent background
(assumed to be $k$-independent).
Both, a small $Z$ and a low Fermi velocity will therefore
make the slope of $n(k)$ very small. However, for the lowest temperature,
$T=0.033t$, one can envisage how the Fermi step
sharpens up at low temperatures.
At $T$$=$$0.1t$, the drop around $k_F$ is still visible, but becomes
very weak. Such a flattening of the Fermi edge discontinuity with
increasing temperature is quite natural.
Inasmuch as the discrete $k$-mesh allows for such
a conclusion, the four $n_f(k)$ for $T\leq 0.1$ all intersect
approximately at $k_F$ - this is what one would expect
if $k_F$ were the Fermi momentum
(on the other hand, the same holds true also for $k_F^0$,
where inspection of $A_f(k,\omega)$ suggests that the drop
is associated with a reduced spectral weight of the
side band). At the intermediate temperature $T$$=$$0.2t$ no more
drop of $n_f(k)$ can be distinguished near $k_F$, and
at $T$$=$$t$ the $f$-distribution becomes completely flat indicating
the more or less
complete localization of the $f$-electrons.\\
Turning to the $c$-electron distribution we first of all note that
all curves intersect at approximately $k_F^0$, and that the
common value of $n_f(k)$ is unity at this momentum. This would
suggest that $k_F^0$ is the Fermi momentum in a Fermi liquid
with $Z$$=$$1$ and for $T\ge 0.5t$ $n_c(k)$ is in fact
nearly identical to that
of free, unhybridized  conduction electrons. 
This changes as the temperature
falls to $T$$=$$0.2t$, where $n_c(k)$ becomes temperature
independent. This would be consistent
with the assumption that the formation of the
heavy band below $T$$=$$0.2t$ the true Fermi surface is shifted
from $k_F^0$ to $k_F$, so that below
this temperature the drop of $n_c(k)$ at $k_F^0$
is due to $k$-dependent hybridization with the heavy $f$-band, rather
than a Fermi surface crossing.\\
Taken together, the data for the single particle spectral function
and the momentum distribution rather clearly suggest a temperature
induced volume collapse of the Fermi surface. At low temperatures
a nearly dispersionless band of low $f$-like spectral weight
crosses the chemical potential near $k_F$, the Fermi momentum for $c$-electrons
and $f$-electrons combined. Thereby in agreement
with analytical work\cite{eos} the heavy band is pinned
to the chemical potential of the Fermi sea of
unhybridized conduction electrons. The momentum distribution
of the $c$-electrons still shows a sharp drop at $k_F^0$\cite{eos,moukuri} 
which becomes $T$-independent at low $T$ and thus cannot develop into
a Fermi edge at $T$$=$$0$. As the temperature increases the heavy band
disappears, whence the Fermi momentum must collapse to $k_F^0$.
Consequently the $f$-electron momentum distribution then shows no more notable
variation near $k_F$, whereas the drop of the $c$-electron
distribution near $k_F^0$ becomes temperature dependent.\\
In summary, we have studied the single particle
spectrum of the one-dimensional Kondo lattice.
A spectacular feature from the technical point of view is that the
notorious minus-sign problem is nearly absent in
this system, so that reliable QMC simulations can be performed even at 
very low temperatures. This opens the possibility to perform detailed 
numerical studies of the nontrivial temperature 
evolution of its electronic structure. Thereby full advantage can be made of 
the special features of QMC, such as the momentum-resolved calculation of all 
dynamical correlation functions.\\
Our data show that despite its one-dimensional character, which would
place the fully interacting system into the universality class of
Luttinger liquids, the system shows a behavior which is quite
reminiscent of the Fermi liquid-like three-dimensional 
Heavy Fermion compounds.
In particular we could find the equivalents of the two characteristic 
temperatures of Heavy Fermions, the Kondo and the coherence temperature.
Both
characteristic temperatures are associated with a distinct change of the
electronic structure near $\mu$: at the lower temperature, which we
associate with the coherence temperature, the `heavy' $f$-like band,
which crosses the Fermi energy and forms the Fermi surface, disappears.
Above this temperature the Fermi surface is $c$-like, and does not
comprise the $f$-electrons any more. Only two dispersionless $f$-like 
sidebands remain at relatively low but finite excitation energy.
Around the second characteristic temperature
these side bands disappear as well, so that
the $c$- and  $f$-electron system
become more or less decoupled.\\
In our previous work on the Kondo insulator\cite{koiso} 
we have also shown that the
temperature where the $f$-electrons `drop out' from the Fermi surface
volume depends sensitively on the $c$-$f$ hybridization $V$
(this is also to be expected on the basis of the impurity 
model\cite{Gunnarson}), and we may expect that the same holds
true also for the metallic case. At a given temperature
different compounds with different $c$-$f$ hybridization,
such as  CeRu$_2$Si$_2$ and CeRu$_2$Ge$_2$, may therefore
be in different regimes, whence their Fermi surface volume
would differ by the number of $f$-electrons.\\
We thank W. Hanke for instructive discussions.
This work was supported by DFN Contract No. TK 598-VA/D03, by BMBF
(05SB8WWA1),
computations were performed at HLRS Stuttgart and HLRZ J\"ulich.
 
\end{multicols}
\end{document}